\documentclass[12pt]{iopart}
\usepackage{graphicx}

\newcommand{\Eqn}[1]{Eq.~(\ref{#1})}     
     
\newcommand{\Sec}[1]{Section~\ref{#1}}     
     
\newcommand{\Fig}[1]{Fig.~\ref{#1}}     
\newcommand{\Figs}[1]{Figs.~\ref{#1}}     
     
\newcommand{\Richerthan}{{\mathcal{P}}_{>}}
\newcommand{\Poorerthan}{{\mathcal{P}}_{<}}
\newcommand{\LN}[1]{\ln \left( #1 \right)}
\newcommand{\aver}[1]{\left< #1 \right>_\pi}

\newcommand{\DD}{\mathcal{D}}
\newcommand{\weight}[1]{\mathcal{W}(#1)}
\newcommand{\tprob}[1]{\mathcal{W}_{#1}}

\begin{document}
\title{Multiplicative Asset Exchange with Arbitrary Return Distributions} 
\author{Cristian F.~Moukarzel$^1$\footnote{email address: cristian@mda.cinvestav.mx}}
\address{$^1$ Depto.\ de F\'\i sica Aplicada, CINVESTAV del IPN,\\
  Av.~Tecnol\'ogico Km 6, 97310 M\'erida, Yucat\'an, M\'exico. }
\date{\today}
\begin{abstract}
  The conservative wealth-exchange process derived from trade
  interactions is modeled as a multiplicative stochastic transference
  of value, where each interaction multiplies the wealth of the
  poorest of the two intervening agents by a random \emph{gain}
  $\eta=(1+\kappa)$, with $\kappa$ a random return.  Analyzing the
  kinetic equation for the wealth distribution $P(w,t)$, general
  properties are derived for arbitrary return distributions
  $\pi(\kappa)$.  If the geometrical average of the gain is larger
  than one, i.e.~if \hbox{$\aver{\ln \eta}>0$}, in the long time limit
  a nontrivial equilibrium wealth distribution $P(w)$ is attained.
  Whenever \hbox{$\aver{\ln \eta}<0$}, on the other hand, \emph{Wealth
    Condensation} occurs, meaning that a single agent gets the whole
  wealth in the long run.  This concentration phenomenon happens even
  if the average return $\aver{\kappa}$ of the poor agent is positive.
  In the stable phase, $P(w)$ behaves as $w^{(T-1)}$ for $w\to 0$, and
  we find $T$ exactly. This exponent is nonzero in the stable phase
  but goes to zero on approach to the condensation interface. The
  exact wealth distribution can be obtained analytically for the
  particular case of \emph{Kelly betting}, and it turns out to be an
  exponential \hbox{$P(w) = e^{-w}$}. We show, however, that our model
  is never reversible, no matter what $\pi(\kappa)$ is.  In the
  condensing phase, the wealth of an agent with relative rank $x$ is
  found to be \hbox{$w(x,t) \sim e^{x t \aver{\ln \eta}}$} for finite
  times $t$.  The wealth distribution is consequently $P(w) \sim 1/w$
  for finite times, while all wealth ends up in the hands of the
  richest agent for large times. Numerical simulations are carried
  out, and found to satisfactorily compare with the above mentioned
  analytic results.
\end{abstract}
%\pacs{}
\maketitle
\section{Multiplicative Trade model}
\label{sec:intro}
The pervasive existence of inequalities in the distribution of wealth
in human societies has puzzled observers since long, but only recently
became a focus of research by physicists~\cite{CYCEWD05}.  The
observation of a power-law distribution for wealth and income in
capitalist societies, originally made by Pareto~\cite{PCDP96}, has
been confirmed and perfected analyzing extensive sets of data that are
nowadays available. These show that the upper five to ten percent of
richest individuals follow a power-law, while the middle to low income
sector of a population follow a Gibbs or lognormal
law~\cite{CCMMIG03,CYCEWD05,YRCSM09}.  The range of typical wealth
variation may be orders of magnitude larger than the natural
variability in individual abilities and capacities, in case one would
attempt to resort to these in order to explain the former. Clearly, it
is of interest to understand what mechanisms may drive the appearance
of such startling differences.  Apart from the fact that wealthier
individuals or entities have, on average, more political power to
influence their social environment to their advantage, thus producing
a self-reinforcing inequality cascade, it is also valid to ask oneself
whether the microscopic mechanisms of wealth production and
redistribution carry within themselves the property of spontaneously
producing inequality.
\\
In a pioneering work~\cite{ANGLE86}, Angle proposed the use of
conservative wealth exchange models in order to explain wealth
inequalities. He envisaged the wealth exchange process as a stochastic
transfer of ``surplus'', in which the looser transfers a random
fraction $\kappa$ of its wealth to the winner. This is nowadays called
the ``looser scheme'' since it is a fraction of the looser's wealth
what is at stake~\cite{IKRWDI98}. If both interacting agents are
assumed to win with the same probability, then such a process avoids
wealth concentration by favoring the poorest agent, because he risks
to loose less than he can win. Aiming to understand the observed
wealth concentration, Angle then argues that the richer agent, because
of the competitive advantage allowed by his larger wealth, usually has
a larger probability of winning in each encounter. Therefore, in the
context of Angle's initial proposal, wealth concentration is a
consequence of an \emph{explicit advantage}, or \emph{edge} favoring
the richer agent~\footnote{An agent is said to have an edge when the
  expectation value of a single trade favors him.}.
\\
Although being richer does provide an advantage in the context of
certain wealth appropriation processes, it is now
recognized~\cite{HFTM02,MGIWCI07} that, from a statistical point of
view, an \emph{explicit} advantage favoring the rich is not a
necessary ingredient for wealth concentration to happen. This is a
remarkable result that only recently has been stressed in the
Econophysics literature. For certain realistic wealth exchange rules
to be discussed in this work, in the long run all wealth may end up in
the hands of just one agent, even if the poor agent has an edge over
the rich one. The key ingredient for this rather counterintuitive
phenomenon is the fact that the amount at stake in each transaction is
a fraction of the poorest agent's
wealth~\cite{IKRWDI98,HFTM02,IGACBR04,SPWATM04,MGIWCI07}, not of the
looser's wealth.  This apparently minor difference in the rules of
these so-called ``poorest scheme'' models profoundly alters the
outcome, as now the poor agent has to be given an explicit advantage
in order to avoid a catastrophic concentration of wealth called Wealth
Condensation. In other words, multiplicative stochastic transfer whose
scale is dictated by the wealth of the poorest intervening agent
implies a ``hidden'' bias in favor of the rich. This is one of the
statistical factors driving wealth concentration.
\\
On the other hand, it can be generally argued that stochastic
multiplicative ``poorest scheme'' transfer rules constitute an
appropriate simple model of the wealth exchange process occurring
during commercial interaction, or trade~\cite{HFTM02,SPWATM04}.
Wealth transfer occurs in a trade operation not because money changes
hands, which is not necessarily always the case, but as a consequence
of the difference in values between the items swapped. This confers
the interaction a clear stochastic character, as none of the agents is
perfectly aware of the true values of the items being interchanged.
Furthermore, it can be argued that the amount at stake must be
proportional to the wealth of the poorest agent, since an interaction
in which the richest agent has the possibility to loose orders of
magnitude more than he can win cannot be considered realistic if one
wishes to reproduce a consensual trade process~\cite{HFTM02,SPWATM04}.
\\
In this work, an analytic and numerical study is presented of wealth
exchange models in which the transference is stochastic,
multiplicative and proportional to the poorest agent's wealth. In each
interaction, the poor agent risks a fraction $\kappa$ of its wealth,
where $\kappa$ is a random variable called the return.  Scafetta,
Picozzi and West~\cite{SPWATM04} have numerically studied a model of
this type in which the return distribution depends on the wealths of
both intervening agents. In order to simplify the derivation of
analytic results, we restrict ourselves to the case in which the
distribution or returns $\pi(\kappa)$ is the same for all pairs of
agents.

\Sec{sec:model} presents the model and a fast heuristic determination
of its condensation interface. In \Sec{sec:kineticequation}, the
kinetic equation for $P(w)$ is introduced and analytic results are
drawn from it. These are compared with numerical results in
\Sec{sec:numerical-results}. \Sec{sec:discussion} presents a
discussion of our results.
\section{Trading with an arbitrary distribution of returns}
\label{sec:model}
\subsection{The model}
Trade interactions are modeled as a process in which wealth is
stochastically transferred between a pair of agents, according to the
following rules. In each transaction a pair of agents is chosen at
random and the poorest one, initially with wealth $w^{poor}$, receives
a gain $ \kappa w^{poor}$, where \hbox{$-1 < \kappa < 1$} is a random
return with distribution $\pi(\kappa)$. The richest agent's wealth
changes by $-\kappa w^{poor}$. The transaction is thus conservative,
and given by
\begin{equation} \label{eq:1}
\left \{
\begin{array}{rcl}
w_{t+1}^{poor} &=&  w_{t}^{poor} + w_{t}^{poor} \kappa_t  \\
w_{t+1}^{rich} &=&  w_{t}^{rich}   - w_{t}^{poor} \kappa_t.
\end{array} \right .
\end{equation}
The condition $|\kappa| < 1$ ensures that both agents have positive
wealth after the trade.
\\
Yard-Sale~\cite{HFTM02,SSMW03,IGACBR04,MGIWCI07} is a particular case
of this process that can be described as a ``bet'' for a fraction $f$
of the wealth of the poorest agent, and where the poorest agent has a
probability $p$ to win. Therefore $\kappa=+f$ with probability $p$ and
$\kappa=-f$ with probability $q=(1-p)$, so \hbox{$\pi^{\hbox{\tiny
      YS}}(\kappa)=p\delta(\kappa-f)+q\delta(\kappa+f)$}.
\subsection{Wealth Condensation}
Depending on $\pi(\kappa)$, long term evolution under rules
(\ref{eq:1}) may give rise either to a stable wealth distribution
$P(w)$ or to wealth condensation~\cite{MGIWCI07}. The surface
consisting of distributions $\pi(\kappa)$ separating these two cases
is called \emph{condensation interface}.  We now derive the location
of this interface as follows: Consider an agent who has become so poor
that, in most subsequent trades he will be the poorest. His own wealth
will thus almost always evolve according to
\begin{equation}
  \label{eq:6}
w_{t+1} =   w_{t} ( 1 + \kappa_t ),
\end{equation}
i.e.~it will undergo a Random Multiplicative Process~\cite{RRMP90}
with \emph{multiplier} $\eta_t=(1+\kappa_t)$ at each timestep. After a
large number $N$ of timesteps, the appropriate central tendency
estimator for its wealth is therefore not the arithmetic average
\begin{equation}
\label{eq:24}
  \aver{w_N} = w_0  \aver{ 1+\kappa }^N,
\end{equation}
but  the geometric average
\begin{equation}
  \label{eq:8}
  e^{\aver{\ln  w_N}} = w_0  \ e^{N \aver{\LN{ 1+\kappa} }},
\end{equation}
Clearly the wealth of a poor agent will diminish steadily if
\hbox{$\aver{\LN{1+\kappa}} < 0$}, in which case there is a sustained
transference of wealth from poorer to richer agents, the system is in
a condensing phase, and the whole wealth ends up in the hands of one
agent in the long run~\cite{MGIWCI07}.  This catastrophic collapse of
the wealth distribution is called \emph{wealth condensation}. By the
heuristic arguments above, the condensation interface is therefore
defined by
\begin{eqnarray}
  \label{eq:17}
  <\ln(1+\kappa)> &=& 0.
\end{eqnarray}
This result will be rederived later in \Sec{sec:cond-interf} by means
of a rigorous analysis of the kinetic equation for this process.
\section{Kinetic Equation Analysis}
\label{sec:kineticequation}
\subsection{Kinetic Equation in the Stationary Limit}
In \ref{sec:kinetic-equation} we show that the equilibrium wealth
distribution $P(w)$ satisfies
\begin{eqnarray}\label{eq:5}
  P(w) = 
\aver{
\frac{P(\frac{w}{1+\kappa})  \Richerthan (\frac{w}{1+\kappa})}
{(1+\kappa)} 
+    \int_0^{\frac{w}{1-\kappa}} dv P(v) P(w+v\kappa) 
},
\end{eqnarray}
where $\Richerthan(w)=\int_w^\infty P(v)dv$ is the fraction of agents
with wealth above $w$, and $\aver{}$ indicates expectation value with
respect to the return distribution $\pi(\kappa)$. The first and second
terms on the right hand side of (\ref{eq:5}) represent the
contributions of exchanges with agents that have a wealth respectively
larger and smaller than $w$.  This equation can be solved exactly
only for special cases that we discuss later in
\Sec{sec:exponential-solution}. In the general case, however, useful
exact results can still be extracted from it, as described next.
\subsection{Small-wealth limit for $P(w)$}
\label{sec:small-wealth-limit}
The small-wealth behavior of the wealth distribution $P(w)$ in the
stable phase can be derived as follows. Assume $P(w) \sim w^{(T-1)}$
for $w\to 0$. Plug this expression into the stationary kinetic
equation (\ref{eq:5}), approximate $\Richerthan(w) \approx 1$ for
small $w$, and notice that the last integral only contributes higher
order terms, to find
\begin{eqnarray} 
\label{eq:10}
\aver{ \frac{1}{ (1+\kappa )^{T}} } = 1.
\end{eqnarray}
This result can be rationalized by referring to Kesten
processes~\cite{LSPLA96,TSTSIV97,SMPA98}, as discussed in
\Sec{sec:discussion}.  Numerical results to be presented later in
\Sec{sec:numerical-results} support the validity of \Eqn{eq:10} in the
limit of small wealth.
\\
Numerical simulation (\Sec{sec:numerical-results}) shows that, using
the value of $T$ resulting from (\ref{eq:10}), the entire wealth
distribution can be \emph{approximated} by a gamma-function
\begin{equation}
  \label{eq:33}
P(w) = a w^{(T-1)} e^{-w/b} ,
\end{equation}
where the normalization conditions on the zeroth and first moments of
$P(w)$ fix $a=\Gamma(T+1)^{T}/\Gamma(T)^{T+1}$, and
$b=\Gamma(T)/\Gamma(T+1)$. Notice, however, that the wealth
distribution (see \Figs{fig:bin.T1T2} and \ref{fig:flat.T1T2}) is not
exactly given by (\ref{eq:33}), except in special cases.
\subsection{Condensation interface}
\label{sec:cond-interf}
Equation (\ref{eq:10}) allows us to determine the location of the
condensation interface by the following rigorous argument. Given that
$P(w) \sim w^{(T-1)}$ for $w \to 0$, the fraction
$\Poorerthan(\epsilon) =\int_{0}^{\epsilon}P(v)dv$ of agents whose
wealth is below an arbitrarily small but finite level $\epsilon$ is
finite for all $T>0$, but diverges as $T \to 0^{+}$. The divergence of
$\Poorerthan(\epsilon)$ indicates that most agents impoverish
absolutely, equivalently that all wealth concentrates in the hands of
a few ones. Therefore, the condensation interface is defined by the
condition $T \to 0^+$. Now rewrite (\ref{eq:10}) as
\hbox{$<e^{-T\ln{(1+\kappa)}}>=1$} and expand it in powers of $T$ to
obtain
% \begin{eqnarray}
% \sum_{r=1}^\infty \frac{(-1)^r T^r <\ln^r(1+\kappa)>}{r} = 0.
% \end{eqnarray}
\begin{eqnarray}
\sum_{r=1}^\infty (-1)^r T^r/r \  <\ln^r(1+\kappa)>/r = 0.
\end{eqnarray}
After eliminating the trivial solution $T=0$, we are left with
\begin{eqnarray}
  <\ln(1+\kappa)>   &=& T \left (    \sum_{r=0}^\infty 
\frac{(-1)^r T^{r} <\ln^{r+2}(1+\kappa)>}{(r+2)} 
  \right ) .
\end{eqnarray}
Therefore, to lowest order, 
% \begin{eqnarray}
%   \label{eq:18}
%   T   &\approx& 
% \frac{2 <\ln(1+\kappa)>}{<\ln^2(1+\kappa)>}.
% \end{eqnarray}
\begin{eqnarray}
  \label{eq:18}
  T   &\approx&   2 <\ln(1+\kappa)>/<\ln^2(1+\kappa)>.
\end{eqnarray}
We thus find that the condensation condition $T\to 0$ amounts to
\hbox{$ \aver{\LN{1+\kappa}} = 0$}, which is the same as \Eqn{eq:17},
derived previously by analyzing the typical behavior of a poor agent's
wealth.
\\
For the case of Yard-Sale, where $\kappa=\pm f$ with probabilities $p$
and \hbox{$q=(1-p)$}, the condensation interface (\ref{eq:17}) is
given by \hbox{$p_c=\LN{1/(1-f)}/\LN{(1+f)/(1-f)}$}, a result that has
been verified numerically~\cite{MGIWCI07}.  Notice that $p_c > 1/2$,
i.e.~the poor has to be given a significant explicit advantage, or
\emph{edge}, in order for condensation not to occur.
\\
For a flat distribution of returns between two limits $a < \kappa <
b$, on the other hand, the critical condition for condensation reads
\begin{eqnarray}
\aver{\LN{\eta}}=\LN{(1+b)^{(1+b)}/(1+a)^{(1+a)} } - (b-a)=0.
\end{eqnarray}
Here again, notice that it is possible to have condensation even when
the average return of the poor agent, which is
$\aver{\kappa}=(a+b)/2$, is positive. 
\subsection{Exponential Solution and Kelly Betting}
\label{sec:exponential-solution}
We now show that, for certain return distributions $\pi(\kappa)$, the
equilibrium wealth distribution is exponential.  For this we replace
$P(w)=e^{-w}$ into (\ref{eq:5}), and get
\begin{equation}
\label{eq:26}
\aver{
(
 e^{-w \frac{1-\kappa}{1+\kappa}}
-e^{-w \frac{1+\kappa}{1-\kappa}}
+1
)
/(1+\kappa) 
}
= 1,
\end{equation}
which is a sufficient condition for the stable wealth distribution to
be exponential.  Particularized to $w=0$, this condition reduces to
\hbox{$\aver{1/(1+\kappa)}=1$}, which is just \Eqn{eq:10} in the case
$T=1$, as appropriate for an exponential distribution. However, notice
that (\ref{eq:26}) is much more restrictive than just (\ref{eq:10})
with $T=1$, because it has to be satisfied for all $w$.
Trivially, if two return distributions $\pi_1(\kappa)$ and
$\pi_2(\kappa)$ satisfy (\ref{eq:26}), so does any normalized linear
combination of them. Therefore, a meaningful approach to solving
(\ref{eq:26}) consists in first finding simple return distributions
which satisfy it, and then building more general ones by linear
combination.  Proposing a binary distribution
\hbox{$\pi(\kappa)=p\delta(\kappa-a)+q\delta(\kappa-b)$}, one finds
that \Eqn{eq:26} is only satisfied if $-a=b=f$ and, additionally,
$p=(1+f)/2$, that is
\begin{equation}
\label{eq:27}
\pi^{(f)}(\kappa) = \frac{1+f}{2} \delta(\kappa - f) 
+  \frac{1-f}{2}\delta(\kappa +f)  
\end{equation}
This return distribution corresponds to Yard-Sale~\cite{MGIWCI07} but
particularized to the case of Kelly betting~\cite{KANI56,RTTKC92},
which fixes $f=2p-1$. Possible implications of this result are
explored in \Sec{sec:discussion}.
\subsection{More general return distributions with exponential solutions}
An arbitrary superposition of return distributions of the form
(\ref{eq:27}) with different values of $f$ will also admit an
exponential solution for $P(w)$. In other words, for any positive
$\weight{f}$ normalized in $[0,1]$ we have that
\begin{equation}
\pi(\kappa) = 
\int_{0}^{1} df \weight{f} \left \{
\frac{1+f}{2} \delta(\kappa - f) +  \frac{1-f}{2}\delta(\kappa +f)  
\right \},
\end{equation}
which is easily integrated to give
\begin{equation}
  \pi(\kappa) = \frac{1 + \kappa}{2}
  \weight{|\kappa|},
\end{equation}
satisfies (\ref{eq:26}), i.e.~gives rise to $P(w)=e^{-w}$ in
equilibrium.
\subsection{Wealths by rank in the condensing phase}
\label{sec:wealths-rank-cond}
Let $w_R$ be the wealths of the $N$ agents ordered by rank $R$, so
that $w_1 > w_2 > \ldots > w_N$.  When an agent with rank $R$
interacts with another agent with rank $S$, we have
\begin{equation}
\label{eq:23}
 w^{(t+1)}_R = \left \{
   \begin{array}{lcr}
     w^{(t)}_R  +\kappa w^{(t)}_R = w^{(t)}_R \eta & \hbox{if} & R>S \\
     w^{(t)}_R  - \kappa w^{(t)}_S  &\hbox{if}  & R<S, 
   \end{array}
\right . 
\end{equation}
Notice that (\ref{eq:23}) is not valid in general, since it disregards
rank changes resulting from interactions. Its validity is restricted
to the case in which agents keep fairly constant ranks, i.e.~there is
no ``social mobility''.  This holds in the condensed phase, but not in
the stable phase.  In the condensed phase, furthermore, one has
$w_S/w_R << 1$ so that the interaction with poorer agents can be
neglected altogether, to write
\begin{equation}
\LN{w^{(t+1)}_R/w^{(t)}_R} \approx \left \{
   \begin{array}{lcr}
    \ln \eta    & \hbox{if} & R>S \\
     0   &\hbox{if}  & R<S. 
   \end{array}
\right . 
\end{equation}
Averaging over $\pi(\kappa)$, over all $(N-1)$ possible choices of
$S$, and defining the relative rank $r=(R-1)/(N-1)$ so that $r=0$
corresponds the richest agent,
\begin{eqnarray}
  \label{eq:22}
  \aver{\LN{w^{(t+1)}_R/w^{(t)}_R }} =   r \aver{\ln \eta} 
  \Rightarrow
  \aver{\ln {w^{(t)}_R}} =   r t \aver{\ln \eta} 
\end{eqnarray}
The typical value of $w(r,t)$ therefore satisfies
\begin{eqnarray}
  \label{eq:20}
  w(r,t) \sim   e^{-rt  \phi},
\end{eqnarray}
where we have defined $\phi= -\aver{\ln \eta} >0$. 
%\begin{trickynorm}
  Normalization for a system of $N$ agents with a total wealth $W$
  then results in
\begin{eqnarray}
  w(r,t) = W \frac{(1- e^{-t\phi/N})}{(1- e^{-t\phi})}    e^{-r  t\phi} 
%\approx W   e^{-r  t\phi} \quad \hbox{for $t \phi >> N$}. 
\end{eqnarray}
%\end{trickynorm}
Now since $r =\Richerthan{(w(r))}$ we have that $P(w) = -1/ (\partial
w(r)/\partial r)$. From (\ref{eq:20}) we thus obtain
\begin{eqnarray}
\label{eq:15}
 P( w ) \sim  \frac{1}{ w}
\end{eqnarray}
The validity of (\ref{eq:20}) and (\ref{eq:15}) is verified for
uniformly distributed returns in \Sec{sec:numerical-results}.
\section{Numerical Results}
\label{sec:numerical-results}
\subsection{Stable phase}
\begin{figure}[!ht]
    \textbf{a)}   \includegraphics[width=0.30\linewidth,angle=270]{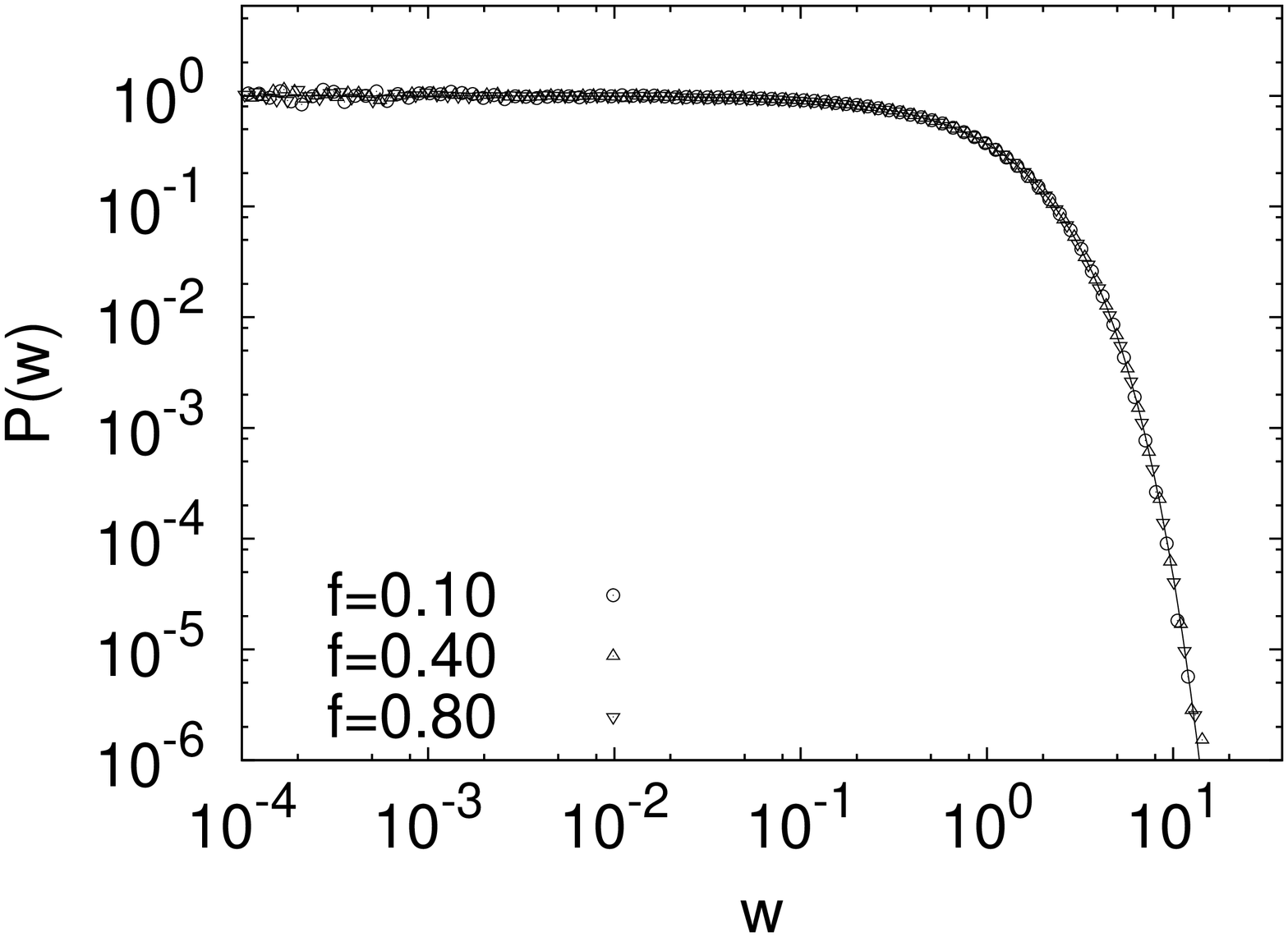}
    \textbf{b)}   \includegraphics[width=0.30\linewidth,angle=270]{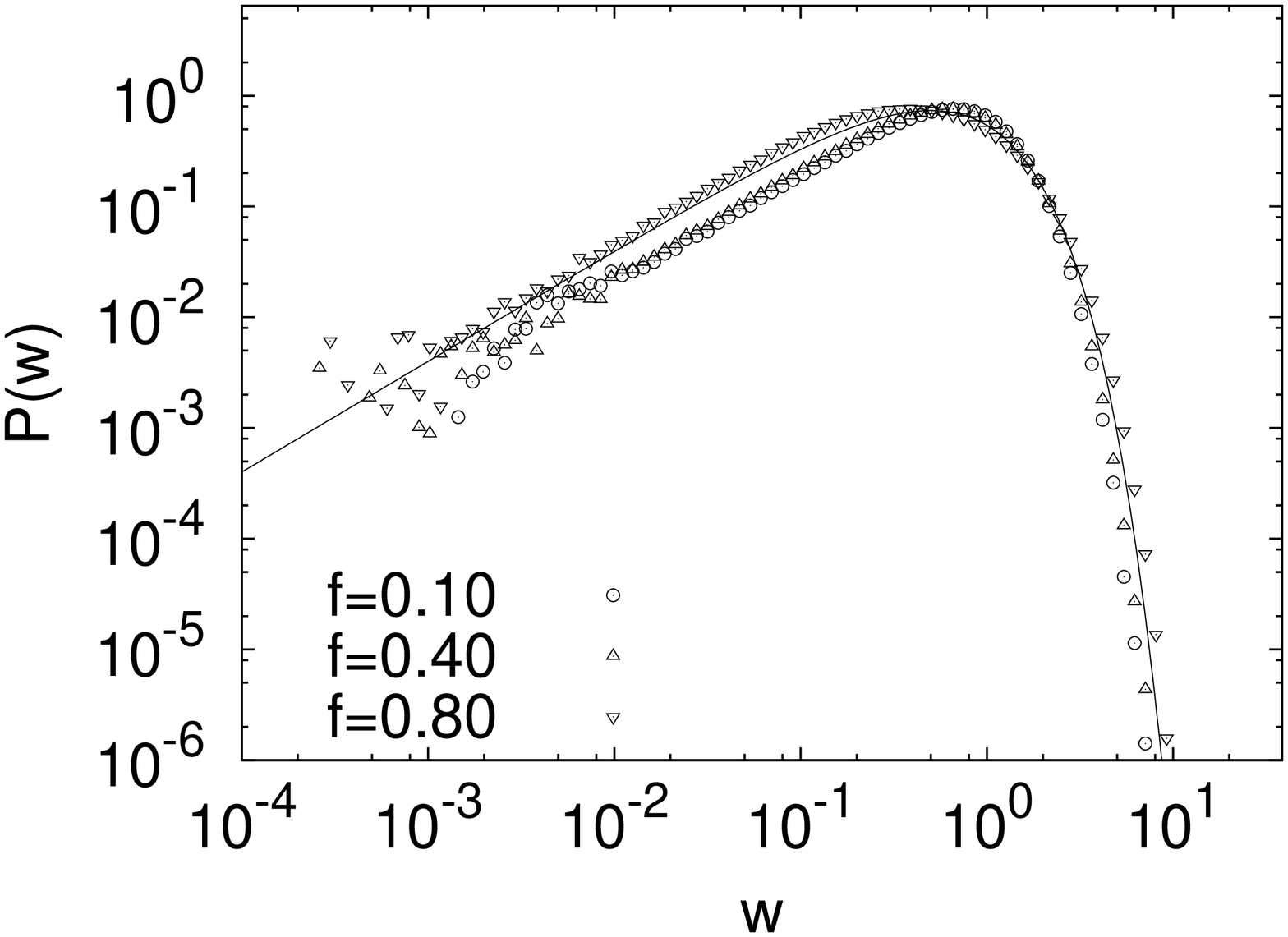}
  \caption{ Wealth distribution for multiplicative exchange with
    binary return distribution ( Yard-Sale ), with $f$ as indicated in
    the figures.  \textbf{a)} $p(f)=(f+1)/2$, which corresponds to
    $T=1$ according to (\ref{eq:35}). The full line is
    $e^{-w}$. \textbf{b)} $p(f)$ satisfies (\ref{eq:36}), for which
    $T=2$ is expected. The full line is a Gamma function (\ref{eq:33})
    with $T=2$.  }
\label{fig:bin.T1T2}
\end{figure}
We first consider the case of Yard-Sale exchange~\cite{MGIWCI07},
i.e. $ \pi(\kappa) = p \delta(\kappa - f) + (1-p) \delta(\kappa +f)$.
\Eqn{eq:10} amounts in this case to
\begin{eqnarray} 
\frac{p}{(1+f)^T}+\frac{q}{(1-f)^T}=1.
\end{eqnarray}
Some particular cases of interest are
\begin{eqnarray}
    \label{eq:34}
    T =0 \quad &\Rightarrow& \quad p = p_c = -\LN{1-f} /
    \LN{\frac{1+f}{1-f}}    \\
    \label{eq:35}
    T =1 \quad &\Rightarrow& \quad  p=\frac{1+f}{2} \\
    \label{eq:36}
    T =2 \quad &\Rightarrow& \quad p=\frac{1}{2}+\frac{3f-f^3}{4}
\end{eqnarray}
\Eqn{eq:34} defines the condensation interface. Its accuracy has been
numerically verified in previous work~\cite{MGIWCI07}.
\Fig{fig:bin.T1T2}a shows wealth distributions for three values of
$f$, where $p(f)$ is given by (\ref{eq:35}), and therefore correspond
to $T=1$, i.e.~$P(w)$ should be constant for $w\to 0$. Notice that all
values of $f$ in this figure satisfy $P(w)=e^{-w}$. This is consistent
with the results derived in \Sec{sec:exponential-solution}, namely
that the wealth distribution is exponential whenever $f=2p-1$ is
satisfied. So in this case, for any pair $(p,f)$ satisfying
(\ref{eq:35}) the wealth distribution is the same.  For $T=2$, the
wealth distribution should approach the origin as $P(w) \sim w$.  This
is verified by considering the data shown in \Fig{fig:bin.T1T2}b,
obtained with $p(f)$ given by (\ref{eq:36}). However, in this case,
notice that the wealth distribution does depend on $f$, i.e.~the
asymptotic exponent $T$ does not determine the whole distribution. A
similar observation holds for the rest of the $(f,p)$ plane: the
equilibrium distribution depends on $f$ on all lines of constant $T$,
except on the line $f=2p-1$, where $T=1$ and $P(w)=e^{-w}$.
\\
\begin{figure}[!ht]
    \textbf{a)}   \includegraphics[width=0.30\linewidth,angle=270]{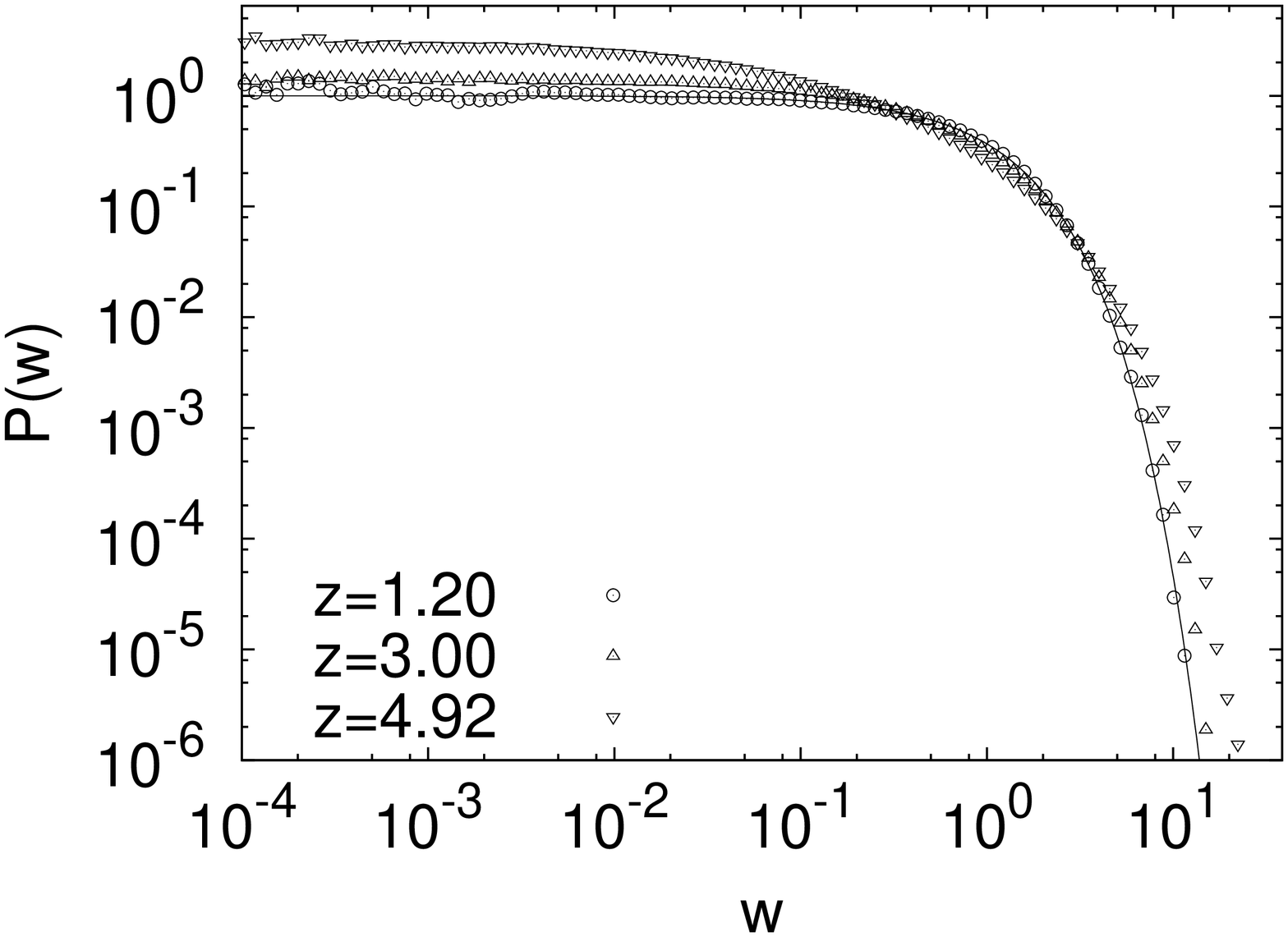}
    \textbf{b)}   \includegraphics[width=0.30\linewidth,angle=270]{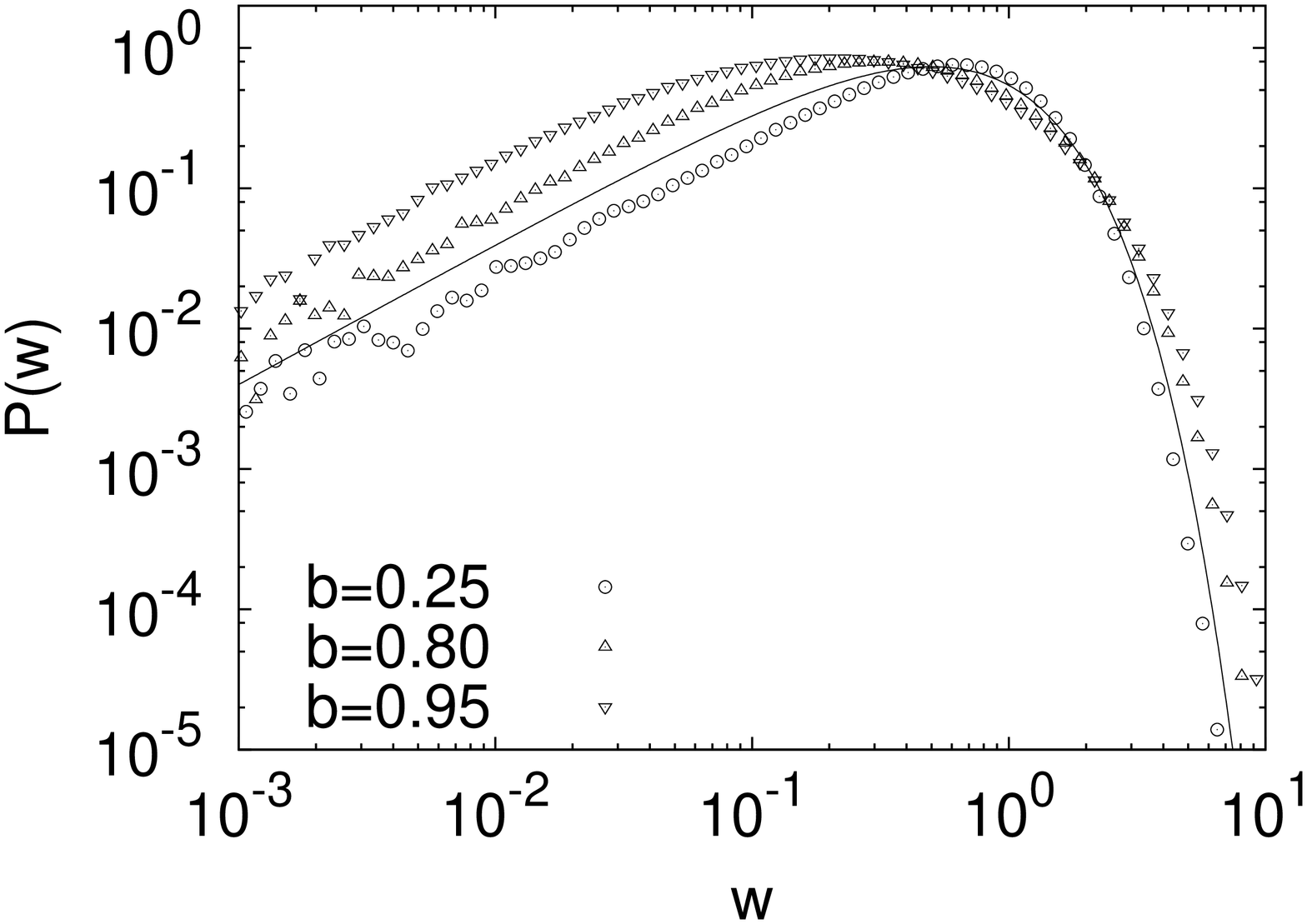}
  \caption{ Wealth distribution for multiplicative exchange with flat
    return distribution between $a$ and $b$. \textbf{a)} $a = \log z
    /(z-1) -1$ and $b = z \log z/(z-1)-1$, with $z$ as indicated in
    the figure. This corresponds to $T=1$, according to \Eqn{eq:39}
    (see text). The full line is $e^{-w}$.  \textbf{b)} Here
    $a=-b/(1+b)$, with $b$ as indicated in the figure. This
    corresponds to $T=2$, according to \Eqn{eq:40}.  The full line is
    a Gamma function (\ref{eq:33}) with $T=2$.  }
\label{fig:flat.T1T2}
\end{figure}
Let us now consider a flat return distribution $\pi(\kappa)=1/(b-a)$
for $a \leq \kappa \leq b$, where $|a|,|b|\leq 1$. This last condition
ensures that the gain $\eta=(1+\kappa)$ lies between zero and two, and
therefore that both agents can always pay.  \Eqn{eq:10} can be worked
out exactly also in this case, and the result is
\begin{eqnarray}
  \label{eq:28}
 \frac{(1+b)^{(1-T)} - (1+a)^{(1-T)}}{(b-a)(1-T)}=1
\end{eqnarray}
We consider the following particular cases
\begin{eqnarray} \label{eq:37}
    \label{eq:38}
    T =0 \quad &\Rightarrow& \quad  \LN{(1+b)^{(1+b)}/(1+a)^{(1+a)}} =b-a 
    \\
    \label{eq:39}
    T =1 \quad &\Rightarrow& \quad  \LN{(1+b)/(1+a)} = (b-a) \\
    \label{eq:40}
    T =2 \quad &\Rightarrow& \quad  (1+b)(1+a)=1.
\end{eqnarray}
The condensation interface has been obtained numerically (not shown)
and found to be in accordance with (\ref{eq:38}).  In the case of
$T=1$, \Eqn{eq:39} can be shown to be equivalent to writing $a = \log
z /(z-1) -1$ and $b = z \log z/(z-1)-1$, where $z=e^{(b-a)}$ is a free
parameter restricted to $1 \leq z \leq 4.9215$. We have
simulated flat return distributions with $a$ and $b$ given by the
expressions above with several values of $z$ (\Fig{fig:flat.T1T2}a),
and found wealth distributions consistent with $T=1$ in all cases,
i.e.~reaching $w=0$ as a constant. Notice that only in the limit $z\to
1$, in which case $a,b \to 0$, the distribution approaches an
exponential. When $T=2$, \Eqn{eq:40} amounts to letting $a=-b/(1+b)$,
within the limits given by $0\leq b \leq 1$. Examples are shown in
\Fig{fig:flat.T1T2}b, where again good accordance with analytic
predictions for the small-wealth exponent is found.
\\
\subsection{Condensed phase}
\begin{figure}[!ht]
    \textbf{a)}   \includegraphics[width=0.30\linewidth,angle=270]{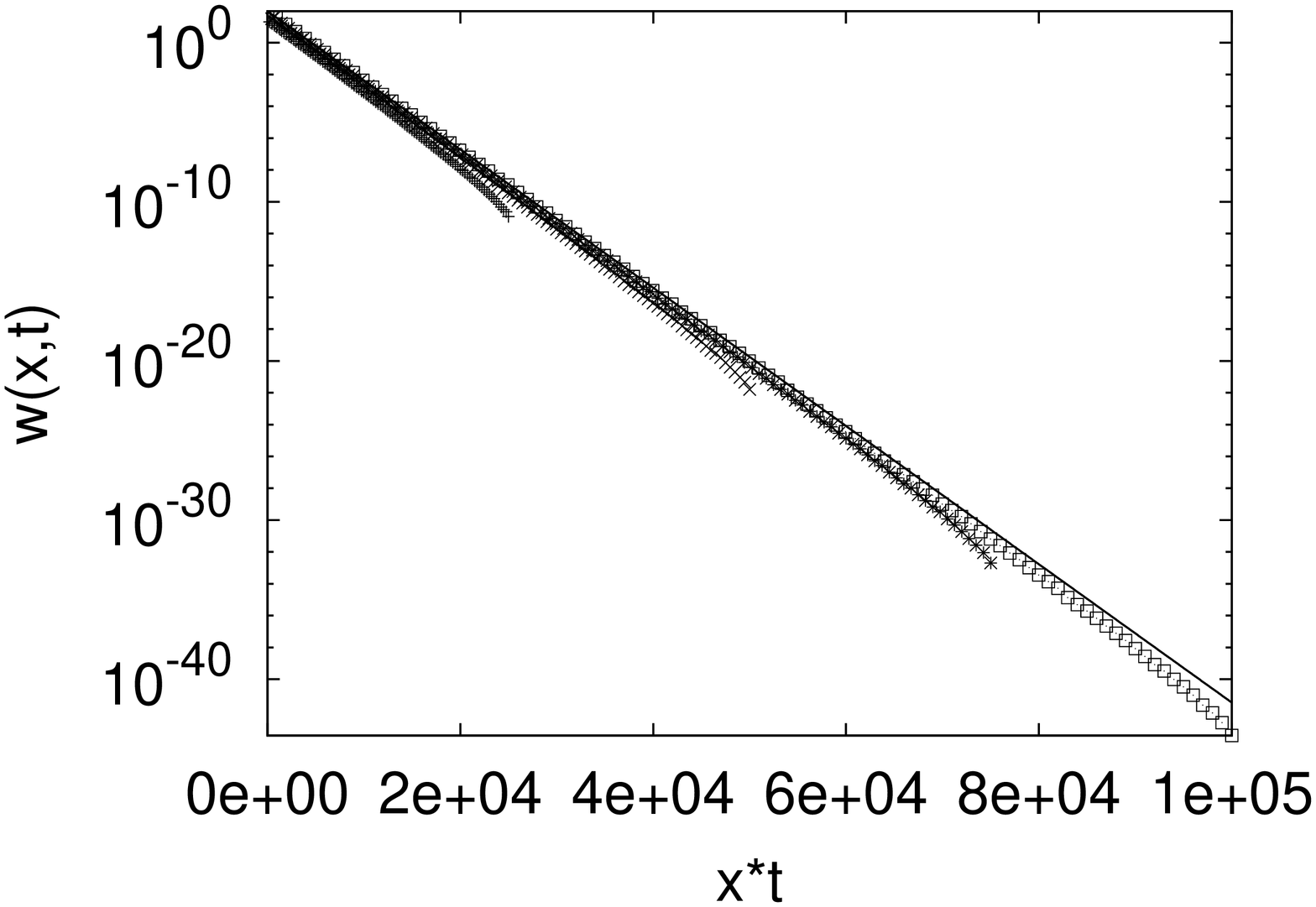}
    \textbf{b)}   \includegraphics[width=0.30\linewidth,angle=270]{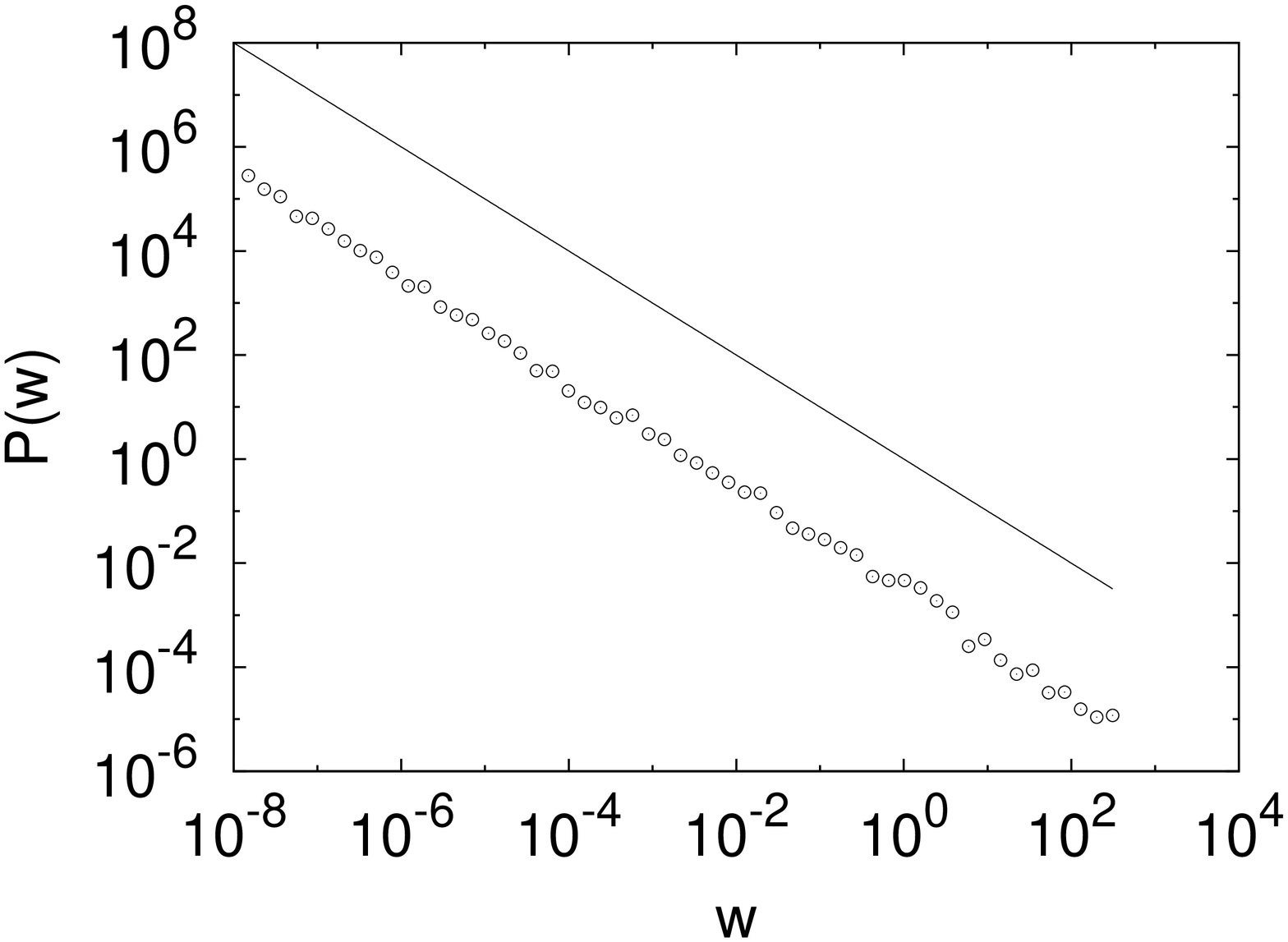}
  \caption{\textbf{a)} Wealth $w(x,t)$ of an agent with relative rank
    $x$, $2.5\times 10^4$ (plusses), $5 \times 10^4$ (crosses),
    $7.5\times 10^{4}$ (asterisks) and $10^5$ (squares) timesteps
    after starting from an egalitarian distribution. The return
    $\kappa$ is distributed uniformly between $-0.01$ and $0.00$, so
    the system is in a condensing state. The full line indicates the
    analytic prediction given by (\ref{eq:20}).  \textbf{b)} Wealth
    distribution $P(w)$ for this case at time $t=10^5$ (not
    normalized). The full line is $1/w$.  }
\label{fig:condensingranked}
\end{figure}

We now verify the approximate expression (\ref{eq:20}), derived in
\Sec{sec:wealths-rank-cond}, for the wealth of an agent with relative
rank $x$ in the condensing state. \Fig{fig:condensingranked}a shows
$w(x,t)$ for returns $\kappa$ distributed uniformly between $-0.1$ and
$+0.1$. Therefore $\aver{\ln \eta}=-1.67\times 10^{-3}$, and the
system is in the condensing state.  The analytic prediction given by
(\ref{eq:20}) is seen in this case to be acceptable, except perhaps
for the very poorest agents. Accordingly, the wealth distribution
$P(w)$ obtained in this condensing state is, as shown in
\Fig{fig:condensingranked}b, consistent with $1/w$ as derived in
\Sec{sec:wealths-rank-cond}.

\section{Discussion}
\label{sec:discussion}
Multiplicative ``poorest-scheme'' asset-exchange models with an
arbitrary return distribution $\pi(\kappa)$ were studied, analyzing
the kinetic equation (\ref{eq:5}).  It was shown that the the whole
system's wealth ``condenses'' onto one agent whenever
$\aver{\LN{1+\kappa} }<0$. Given that $e^{\aver{\ln (1+\kappa)}} \leq
e^{\ln \aver{(1+\kappa)}} =\aver{(1+\kappa)}$, it is possible to have
$\aver{\ln (1+\kappa)}<0$, and therefore wealth condensation, even in
cases in which the average return $\aver{\kappa}$ is positive. But
having a positive average return means, according to (\ref{eq:24}),
that the expectation value of a poor agent's wealth \emph{grows
  exponentially} in time.  The apparent paradox suggested by the fact
that poor agents loose wealth steadily despite their average return
being positive is solved by recognizing that the expectation value is
not an appropriate central-tendency estimator when considering
multiplicative processes~\cite{RRMP90}. In other words, while the
exponential growth indicated by (\ref{eq:24}) would only be realized
after averaging over an enormously large number of repetitions of the
multiplicative process, the typical outcome of one realization follows
the geometric average (\ref{eq:8}), which, in the condensing phase,
decreases exponentially fast in time.
\\
Analyzing the kinetic equation, the small-wealth exponent $T$ of
$P(w)$, was found to be given by \Eqn{eq:10}. This result can be
understood in the context of Kesten
processes~\cite{LSPLA96,TSTSIV97,SMPA98}, as follows. In the stable
phase, where $\aver{\LN{1+\kappa}}>0$, the inverse wealth $z=w^{-1}$
of a poor agent undergoes a \emph{contractive} multiplicative process
\hbox{$z \to z \xi = z/(1+\kappa)$} due to its interaction with richer
agents, with a small additive noise term given by its almost
negligible interaction with poorer agents, as $z \to z +\delta
z$. Because in the stable phase $\aver{\kappa}>0$, it follows that
$\aver{\delta z}> 0$.  The theory of Kesten processes then ensures
that $P(z)$ has a power-law right tail of the form $z^{-(1+T)}$ with
$T$ satisfying $\aver{\xi^T}=1$. By a simple change of variables, our
result (\ref{eq:10}) then follows for $P(w)$.
\\
In \Sec{sec:exponential-solution} it was shown that the wealth
distribution is exactly exponential when $\pi(\kappa)$ corresponds to
Kelly betting~\cite{KANI56,RTTKC92}.  Kelly betting is a gambling
strategy devised to maximize the long-time rate of growth of a
bettor's capital when faced with a set of risky choices. In its
simplest inception, one considers a gambler who is given a choice
between a single risky asset (a bet) and a riskless asset
(e.g.~deciding not to bet). Assuming a bet that pays double or
nothing, the gambler doubles its stake or looses it altogether,
respectively with probabilities $p$ and $q$.  The bettor can decide
the (fixed) fraction $fw$ of his wealth that will be risked at each
time step, while the rest $(1-f)w$ is kept in the riskless asset
(e.g.~cash). His total wealth thus evolves as \hbox{$w\to ( (1-f)w +
  2fw)=(1+f)w$} if the bet is won, and as $w\to (1-f) w$ if the bet is
lost. In other words, the gambler's \emph{gain} is $\eta=(1+f)$ with
probability $p$, and $\eta=(1-f)$ with probability $q$. This is
exactly the way in which the wealth of the poorest agent evolves in
Yard-Sale. The only difference is the fact that, in the betting
optimization problem, $f$ and $p$ are not independent parameters,
since the gambler wishes to find the most profitable $f$ for a given
$p$.
\\
The average gain is $\aver{\eta}=(1+f)p+(1-f)q=1+(2p-1)f$, which is
larger than one whenever $p>1/2$, i.e~whenever the \emph{edge}
$(2p-1)$ is positive. If the bettor were to maximize $\aver{\eta}$,
the recommended strategy would then be choosing $f$ as large as
possible~\footnote{ In this context, $f>1$ may be acceptable and would
  mean that the gambler lends money from the riskless asset for
  gambling. }. However, this approach is doomed to fail in the long
run, since sooner or later a loosing bet would produce his absolute
ruin. Kelly then proposes that the most profitable strategy \emph{in
  the long run} consists in using the value of $f$ maximizing the
average growth rate $G$ of the gambler's wealth, defined by $w(t) \sim
e^{G t}$. The average growth rate is then given by the average
logarithmic gain $\aver{\LN{\eta}}$, since a random multiplicative
process typically follows its geometric average. Maximization of
\hbox{$\aver{\LN{\eta}}=p \LN{1+f}+q\LN{1-f}$} with respect to $f$
then results in $f^*=(2p-1)$, which is the recommendation of Kelly
theory for this problem. Equivalently, $p=(1+f)/2$ and $q=(1-f)/2$,
from which we can recognize that (\ref{eq:27}) describes Yard-Sale
exchange in the case of Kelly betting.
\\
From the discussion above, $f^*$ is the fraction at stake that
produces the fastest growth in the wealth of a poor agent, for a given
$p$. However, notice that this is not the value of $f$ that produces
the least density of poor agents in Yard-Sale for a given $p$. The
density of poor agents in Yard-Sale in equilibrium is minimum in the
limit $f\to 0$, for any fixed $p$. This conclusion may be reached by
noticing that the fraction of poor agents is smaller the larger $T$ is
and analyzing~(\ref{eq:10}) particularized to Yard-Sale, in the limit
$T\to \infty$.
\\
The link between Kelly betting and an exponential wealth distribution
in these exchange models is intriguing. The Kelly strategy can be
restated in the language of information theory as a way to maximize
the rate of transfer of information over a noisy
channel~\cite{KANI56}. On the other hand, the exponential distribution
$P(w)=e^{-w}$ maximizes the entropy $S=-\int P(w) \LN{P(w)} dw$
subject to the constraints of constant total wealth $\int wP(w) dw=1$
and number of players $\int P(w) dw=1$.  Of course, there is in
principle no logical relation between extremization of an entropy
transfer rate, and maximizing the total entropy in equilibrium,
however the connection seems worth analyzing.
\\
Because the dynamics is conservative, if the microscopic exchange
rules are reversible it can be shown~\cite{DYSMO00} that the
equilibrium wealth distribution has to be exponential. However, the
converse is of course not true, i.e.~the existence of a stable
exponential solution does not imply reversibility. In fact,
multiplicative exchange rules of the type discussed here usually
violate reversibility, since the role of both agents is clearly
different. Nevertheless, it could in principle be the case that, for
some specific return distributions, the general exchange rules
considered here were reversible. This in turn would provide a clearcut
explanation for the appearance of an exponential solution.  However,
in \ref{sec:reversibility} it is shown that this is not the case,
i.e.~reversibility is not satisfied for exchange rules of the general
type (\ref{eq:1}), no matter what the return distribution
$\pi(\kappa)$ is.  Therefore, the wealth distribution in equilibrium
is exponential for return distributions of the form (\ref{eq:27}), not
because of reversibility, but because of accidental cancellation of
asymmetries in the transition rules.  
\ack 

Juan Manuel V\'azquez-Montejo participated in the early phases of this
work. I thank Marcio Argollo De Menezes for help with the programming.
\appendix
\section{Kinetic Equation}
\label{sec:kinetic-equation}
Consider an interaction between two agents $Y$ and $Z$ with initial
wealths $y$ and $z$, assuming without loss of generality $y<z$.  The
interaction processes contributing to $\dot{P}(x)$ are those in which
$x$ is the wealth of one of the two interacting agents, either before
or after the interaction. Adopting the shorthand notation \hbox{$\DD
  \kappa= \pi(\kappa) d\kappa$}, \hbox{$\DD y= P(y) d y$}, and
\hbox{$\DD z= P(z) d z$}, one has:
\begin{enumerate}
\item The poorest agent $Y$ has a return $\kappa$, its wealth thus
  becoming $y(1+\kappa) = x$. This contributes with $\DD \kappa \DD y
  \DD z \theta(z-y) \delta(x-y(1+\kappa))$.
\item The richest agent $Z$ takes part in an exchange where the
  poorest agent $Y$ has return $\kappa$. $Z$'s wealth becomes \hbox{$z
    -\kappa y =x$}. The contribution of this process is $\DD \kappa
  \DD y \DD z \theta(z-y) \delta(x-z+y \kappa)$.
\item Either agent has wealth $x$ before the interaction, but not
  after it, resulting in a contribution $ - \DD \kappa \DD y \DD z
  \theta(z-y) (\delta(x-y) + \delta(x-z))$.
\end{enumerate}
If $r$ is the probability per unit time of a trade, the time
derivative of $P(x)$ is then given by
\begin{eqnarray} \label{eq:2}
r^{-1}  \dot{P}(x)  &=&  
\int \DD \kappa   \DD y \DD z  \theta(z-y) 
\left \{ 
 \delta(x-y(1+\kappa)) + \delta(x-z+y\kappa)  
\right .
\nonumber \\ &&
\left .
-  \delta(x-z) -\delta (x-y)
\right \}.
\end{eqnarray}
Letting $\Richerthan(x)=\int_x^\infty P(z)dz$, we are left with 
\begin{eqnarray}
\label{eq:29}
&&  \frac{1}{r}\dot P(x) =    -P(x)  + 
\nonumber \\ &&
\int_{-1}^{1} \DD \kappa   \left \{  
\frac{P(\frac{x}{1+\kappa})  \Richerthan (\frac{x}{1+\kappa})}
{1+\kappa} +
    \int_0^{\frac{x}{1-\kappa}} dy P(y) P(x+y\kappa) 
  \right \}
\end{eqnarray}
It can be seen that this equation conserves the zeroth- and
first-moments of $P(w)$, i.e.~number of agents and total wealth are
conserved.
\section{Reversibility}
\label{sec:reversibility}
We wish to determine whether the exchange rules considered in this
work satisfy reversibility for some return distribution
$\pi(\kappa)$. For this purpose we have to write the kinetic equation
in the general form
\begin{eqnarray}
  \dot P(x) = \int dz d \rho    
  \left \{
    P(x-\rho) P(z) 
    \tprob{ (x-\rho, z) \to (x, z-\rho)} 
  \right . 
  \nonumber \\
  \left .
    -P(x) P(z-\rho)  
    \tprob{ (x, z-\rho) \to (x-\rho, z)} 
  \right \},
\end{eqnarray}
and check whether \hbox{$\tprob{ (x-\rho, z) \to (x, z-\rho)}=\tprob{
    (x, z-\rho) \to (x-\rho, z)}$} is satisfied.  A lengthy but
straightforward calculation shows that
\begin{eqnarray}
\tprob{ (x-\rho, z) \to (x, z-\rho)} = 
&\theta&(z - (x-\rho)) \frac{1}{x-\rho} \pi(\frac{\rho}{(x-\rho)}) + 
\nonumber \\ 
&\theta&((x-\rho) - z) \frac{1}{z} \pi(-\frac{\rho}{z}).
\nonumber \\ 
\tprob{ (x, z-\rho) \to (x-\rho, z)} = 
&\theta&((z-\rho)-x  ) \frac{1}{x} \pi(-\frac{\rho}{x}) +
\nonumber \\ 
&\theta&(x- (z  -\rho) )
\frac{z}{(z-\rho)^2}    \pi( \frac{\rho}{(z-\rho)}).
\end{eqnarray}
If reversibility holds, the following conditions must then be met:
\begin{equation}
\left \{  
\begin{array}{lcrcll}
% Case 1
z>x +|\rho|, &\Rightarrow& 
\frac{1}{x-\rho} \pi(\frac{\rho}{(x-\rho)}) &=&
\frac{1}{x} \pi(-\frac{\rho}{x}) ,
 \\ \\ 
% Case 2
|z-x| < \rho,
&\Rightarrow&
\frac{1}{x-\rho} \pi(\frac{\rho}{(x-\rho)})
&=&
\frac{z}{(z-\rho)^2}  \pi( \frac{\rho}{(z-\rho)}),
 \\ \\ 
% Case 3
\rho < |z-x|,
&\Rightarrow&
\frac{1}{z} \pi(\frac{-\rho}{z})
&=&
\frac{1}{x} \pi(-\frac{\rho}{x}) ,
 \\ \\ 
% Case 4
 z < x-|\rho|,
&\Rightarrow&  
\frac{1}{z} \pi(\frac{-\rho}{z})
&=&
\frac{z}{(z-\rho)^2}  \pi( \frac{\rho}{(z-\rho)}).
\end{array}
\right .
\end{equation}
The first condition is independent of $z$, so it has to hold for all
$x$ and $\rho$. By calling $\rho/(x-\rho)=\kappa$ and after some
manipulation, this condition reads
\begin{eqnarray}
\pi(-\kappa) = \frac{1}{1-\kappa} \pi(\frac{\kappa}{1-\kappa}).
\end{eqnarray}
Similar manipulation of the fourth case gives
\begin{eqnarray}
\pi(-\kappa) = \frac{1}{(1-\kappa)^2} \pi(\frac{\kappa}{1-\kappa}).
\end{eqnarray}
These are only compatible with each other if $\kappa=0$. Therefore the
exchange rules discussed in this work are never reversible.  

\end{document}